**Prediction of kinase inhibitor response using activity profiling, *in-vitro* screening, and elastic net regression.**

T. Tran,[1] E. Ong,[2] A. P. Hodges,[1] G. Paternostro,[1,2] and C. Piermarocchi[2,3]

[1]*Sanford Burnham Institute for Medical Research, La Jolla, CA 92037*

[2]*Salgomed Inc., Del Mar, CA 92014*

[3]*Department of Physics and Astronomy, Michigan State University, East Lansing MI 48824*

**Abstract:**

**Background:** Many kinase inhibitors have been approved as cancer therapies. Recently, libraries of kinase inhibitors have been extensively profiled, thus providing a map of the strength of action of each compound on a large number of its targets. These profiled libraries define drug-kinase networks that can predict the effectiveness of new untested drugs and elucidate the role played by specific kinases in different cellular systems. Predictions of drug effectiveness based on a comprehensive network model of cellular signalling are difficult, due to our partial knowledge of the complex biological processes downstream of the targeted kinases.

**Results:** We have developed the Kinase Inhibitors Elastic Net (KIEN) method, which integrates information contained in drug-kinase networks with *in vitro* screening. The method uses the *in vitro* cell response of single drugs and drug pair combinations as a training set to build linear and nonlinear regression models. Besides predicting the effectiveness of untested drugs, the method identifies sets of kinases that are statistically associated to drug sensitivity in a given cell line. We compare different versions of the method, which is based on a regression technique known as *elastic net*. Data from two-drug combinations leads to predictive models, and predictivity can be improved by applying logarithmic transformation to the data. The method is applied to the A549 lung cancer cell line. A pathway enrichment analysis of the set of kinases identified by the method shows that axon guidance, activation of Rac, and semaphorin interactions pathways are associated to a selective response to therapeutic intervention in this cell line.

**Conclusions:** We have proposed an integrated experimental and computational methodology that identifies the role of specific kinases in the drug response of a given cell line. The method will facilitate the design of new kinase inhibitors and the development of therapeutic interventions with combinations of many inhibitors.

**1. Background**

The important role of kinases in cancer biology[1] has spurred a considerable effort towards the synthesis of libraries of fully profiled kinase inhibitors, providing a map of the strength of each compound on a large number of its potential targets.[2-4] In particular, a recently published dataset has profiled several hundred kinase inhibitors using a panel of more than 300 kinases.[4] These profiled libraries define a network of interactions between drugs and their kinase targets,[5] and represent a valuable resource for the development of new therapies. In this paper, we introduce a computational method using information present in profiled libraries and *in vitro* cell response to predict the response of untested drugs. Besides making prediction, the method identifies critical kinase targets and pathways that are statistically associated to drug sensitivity in a given cell line.

Statistical inference and regression methods in conjunction with gene expression or mutations have been used to identify specific biomarkers associated with an increased sensitivity/resistance to drugs. For instance, the sensitivity to PARP inhibitors of Ewing's sarcoma cells with mutations in the EWS gene and to MEK inhibitors in NRAS-mutant cell lines with AHR expression have been predicted using analysis of variance and the elastic net method[6] and then experimentally validated.[7,8] In these analyses, the statistical variable associated to drugs was represented by the half maximal inhibitory concentration ($IC_{50}$) in different cell lines. However, besides the $IC_{50}$, many other types of information characterize chemical compounds, and these other types of information can be used to enhance the statistical analyses and improve the accuracy of predictions. For instance, a method to predict drugs sensitivity in cell lines based on the integration of genomic data with molecular physico-chemical descriptors of the drugs has been recently proposed.[9] The residual activity of kinases after interaction with a compound is another quantity that can be useful. Kinase profiling, patient genetic profiles, and sensitivity of primary leukemia patient samples to kinase inhibitors have been recently used by Tyner *et al.*[10] to identify functionally important kinase targets and clarify kinase pathway dependence in cancer.

In this paper, the residual activity of kinases upon drug interaction will be used as predictors of the cellular response for *in vitro* experiments and then integrated in a regression method known as elastic net.[6] This regression method reduces the number of predictors to a minimum set, providing a clear picture of the kinases involved in the sensitive response of cell lines. As training data sets for the regression we will use a primary and a secondary screen corresponding to single-drug response and response to combinations of two drugs. The method based on two-drug combinations is particularly important due to the current interest in the development of methods for the discovery of therapies based on drug combinations.[11] Moreover, the two-drug screening exhibits a broader distribution in the response and provides the models with a good level of predictability. In fact, the model based only on single drug response did not pass the statistical cross-validation test.

We are applying this Kinase Inhibitor Elastic Net (KIEN) method to predict the cell viability of a lung cancer cell line (A549) and a normal fibroblast cell line (IMR-90) after drug treatment. We found that the regression can be improved through a logarithmic transformation on the data. Using the results of the regression, we identified a set of kinases that are strongly associated to a response in A549 that is selective with respect to IMR-90. A pathway-based enrichment test was done using the results from this kinase analysis. Ten pathways from Reactome[12] were identified as significant using this set of kinases, including axonal guidance and related semaphorin interactions as top hits.

This paper is organized as follows: We first present in section 2.1 the experimental results of the primary and secondary *in vitro* screening corresponding to single drugs and two-drug combinations. These results are analysed in terms of Pearson's correlation with kinase activity in section 2.2. This simple correlation analysis gives a first glance of the kinases that are statistically associated to a significant change in the viability of cancer and normal cell lines. In section 2.3, we introduce the elastic net approach and we present the results of a leave-one-out cross validation for predictions on single and pairs of drugs. We also present in this section the results obtained using the logarithmic transformation on the variables and a pathway enrichment analysis using Reactome.[12] The discussion of the results is in Section 3, conclusions in Section 4, and Materials and Methods in Section 5.

## 2. Results

### *2.1 In vitro screen of the kinase inhibitor library*

Our methodology begins with the experimental screening of single and pairwise drug response. The 244 kinase inhibitors (KIs) of our drug library were screened at 1000nM individually and the treatment lasted for 72 hours. To quantify a selective response of a cancer cell line with respect to a control normal cell line, we define the selectivity $S$ of a single drug or drug combination as

$$S = \frac{v_N}{v_C}$$

where $0 \leq v_{N(C)} \leq 1$ refers to the viability of normal (N) and cancer (C) cell lines after treatment. From the screening of the 244 KIs, the top hit was PDK1/Akt1/Flt3 Dual Pathway Inhibitor (CAS # 331253-86-2) as ranked by selectivity (Figure 1). For the secondary screen we used the PDK1/Akt1/Flt3 Dual Pathway Inhibitor as the starting point and combined this compound in combination with the other KIs; PDK1/Akt1/Flt3 Dual Pathway Inhibitor was paired with one other KIs for a two-drug pair screen. The dose of PDK1/Akt1/Flt3 Dual Pathway Inhibitor was optimized to ensure proper dosing range. We found its most selective dose to be 125nM (S=10.1± 0.2), while maintaining the normal cell line IMR-90's viability >90% (Figure 2). We used the most selective dose at 125nM and the other 243 KIs at 1000nM for the secondary screen. The secondary screen showed that when PDK1/Akt1/Flt3 Dual Pathway Inhibitor was paired with another KI, the drug pair presented synergistic effects. The resulting top hit from the secondary screen of the library was Alsterpaullone 2-cyanoethyl (CAS # 852529-97-0) with the selectivity of kinase inhibitor drug pairs with S= 6.14 (Figure 3).

### 2.2 Analysis of correlations

In our second step, we analyzed the Pearson's correlation of the primary and secondary screening with a published dataset[4] containing target profiles for 140 kinase inhibitors. Therefore, even though we had a library of 244 KIs in the experimental screening, we were limited to utilizing 140 KIs for the analysis. For each inhibitor, the dataset provides the residual activity ($0 \leq A \leq 1$) of 291 kinases after drug treatment. This quantity is a measure of the strength of inhibition of a drug on each kinase.

For each kinase $k$, we calculate the Pearson's correlation, $C_k$, between the selectivity $S_i$ and the activities $A_{k,i}$, with $i \in \{1, \dots, M\}$ labeling the single drug or drug pair in the set. For drug pairs, the activity is estimated as a product of the residual activities of the two drugs. The kinases are then ranked based on the *p*-value of their correlation with selectivity, and we calculate the False Discovery Rate (FDR) adjusted *p* value.[13] The list of kinases mostly correlated to the selectivity from the primary and secondary screen are listed in Table 1. A similar analysis was done by calculating the correlation between the normal or cancer cell viability $v_i$ and the activities. The results for the top kinase-viability correlations for the primary and secondary screen are shown in Table 2. Note that, although significant correlations are found in both cases, the secondary screen gives stronger correlations than the primary screen.

### 2.3 Elastic Net regression

We build a regression model that predicts the response of a cell line to a drug or drug combination *i*. The response we predict is the normal and cancer cell viability, from which the selectivity can be derived. For this purpose, we define a regression problem in which we use the residual activity of the kinase $k$ under the effect of drug $i$, which we indicate as $A_{k,i}$, as predictors of the viability. The response can be written as

$$v_i = \beta_0 + \beta_1 A_{1,i} + \cdots + \beta_p A_{p,i}. \qquad (1)$$

A fitting procedure based on a training set of measurements produces the coefficients $(\beta_0, \beta_1, \ldots, \beta_p)$. Equation (1) can then be used to predict the viability of a new drug that has not been tested, but of which the profiling information is available. Note that we are integrating two different types of experimental data: kinase profiling data is obtained through enzymatic assays that probe directly the interaction between drug and kinases, while the *in vitro* cell response data is the result of complex signaling that involves many pathways downstream of the affected kinases. The coefficients $\beta_k$ can be seen as a measure of the sensitivity of a given cell line due to alterations in the activity of kinase $k$.

It is well known that the least square method does not perform well in the case of linear regression with many predictors. In our case, we would like to use a database of drugs that have been profiled on about 300 kinases. However, it would be desirable to select and keep in the final model a minimal set of the kinases that perform well as predictors and provide a simple model, useful to gain biological insight. The lasso technique[14] is a powerful method to reduce the number of predictors by imposing a penalty on the regression coefficients. However, in the presence of a group of kinase predictors with strong mutual correlation, the lasso could select only one kinase predictor from the group while missing the others. To prevent this problem, our method uses the elastic net approach. This method incorporates the lasso penalty as well as a ridge penalty to keep the regression coefficients small without completely removing them.[6] The weights of the ridge and lasso penalties in the least square procedure can be optimized for best performance of the method.

We show in Figure 4 (a) and (b) the results of a leave one out cross validation (LOOCV) method for the primary (a) and secondary screen (b). For each of the 140 drugs, we apply the elastic net method using the remaining 139 drugs and then we compare the result to the measured value. This cross validation method is a particular case of the more general $k$-fold cross validation procedure in which $k$ is equal to the size of the training set.[15] The cross LOOCV shows that the information contained in the primary screen is not sufficient to define a predictive model. The fact that some kinases in Tables 1 and 2 show some significant correlation with the response when considered individually is in general not a sufficient condition for defining a predictive multiple regression model. On the other hand, the secondary screen is able to reproduce the viability of many drugs, especially the ones with the stronger effect on both cell lines. Overall, the data from the secondary screen presents a much broader distribution with a tail representing a few drug combinations particularly effective. The regression works better in identifying these highly effective pairwise combinations and the relative ranking of their strengths. Data is not particularly informative for drugs and drug combinations that are not effective, which concentrate in the neighborhood of $\sim 1$ .

Data transformations can represent a powerful strategy to improve regression. We applied a logarithmic transformation, which is consistent with the hypothesis of an independent action on the different kinases on the total viability. We write the viability as

$$v_i = e^{\beta_0} (A_{1,i})^{\beta_1} \cdot (A_{2,i})^{\beta_2} \cdot \ldots \cdot (A_{p,i})^{\beta_p} . \qquad (2)$$

By applying a *log* transformation on both sides of Eq. (2) we reduce the problem to a linear regression, to which the elastic net strategy can be applied. We show in Figure 5 the results of the LOOCV for the primary and secondary screen using the logarithmic data transformation. As in the linear case, we find that the method fails the cross validation procedure if we use data from the primary screen, while the secondary screen with log transformed data gives better $R^2$.

In addition to a regression model that can be used to predict the efficacy of drugs that have not been tested, the $\beta_i$ coefficients can be used to rank kinases in terms of their relevance in the regression. These coefficients therefore identify the kinases whose inhibition is associated to a decrease in the cell viability. A ranking based on the differential $\beta_i^C - \beta_i^N$, where the index $N$ and $C$ identify the regression model of the cancer and normal cells, gives insight on specific pathways important for a selective response of cancer cells. Table 3 show a list of kinases ranked in terms of $|\beta_i^C - \beta_i^N|$, where the coefficients have been obtained using the logarithmic data transformation on the secondary screen, which, as seen above, is the best performing method according to our cross validation.

In order to test whether selected pathways were significantly enriched for the identified kinase genes in Table 3, a pathway-based enrichment analysis was conducted using the results from the kinase analysis and Fisher exact tests. Ten pathways from Reactome were identified as significant (p<0.05) using this kinase list, including axon guidance, activation of Rac, and semaphorin interactions as top hits (Table 4).

## 3. Discussion

Drug-kinase profiling represents a controller-target network[5] that when combined with *in vitro* testing, can be used in regression models to predict response and identify pathways statistically associated to drug sensitivity. Network methods in biology are often based on the analysis of large datasets from high-throughput experiments. An example is given by gene regulatory networks, which presents many challenges either when restricted to a homogeneous set of data[16,17] or when it includes different classes of data.[18-21] In our KIEN method, information from the drug-target network and experimental query of the biological system are integrated. The goal is not a reconstruction of a regulatory network, but we wish to identify a set of kinases linked to a therapeutic response in a given cell line. In order to establish associations, the system has to be perturbed by the use of kinase inhibitor drugs. The response to these drugs or drug combinations provides a training set that when combined with the profiling, can lead to predictions.

The Elastic Net method is one of the most widely used regularization techniques. Regularization techniques are used in statistical and machine learning models to achieve an optimal tradeoff between accuracy and simplicity. Simplicity makes a model less prone to overfitting and more likely to generalize. In our analysis, we found that the elastic net regressions based on single drug responses were not successful, while drug pair data provided statistically significant predictions. A possible explanation for this finding is the following: single drugs might be less able to overcome the robustness of biological networks. The phenotypic signal is therefore blunted and not easily measured. If a second drug is added, any compensatory capacity is already stretched and the effects from the inhibition of each kinase can be seen more clearly. Using data from drug pairs, we found that noise can be better filtered out and stronger statistical associations between kinases and therapeutic response are revealed. Clearly, if a different training set of more effective drugs were used in the primary screen, it is likely that also single drug *in vitro* response would have given a significant predictive model.

Among the top three pathways shown in Table 4 are activation of Rac and Semaphorin interactions. Rac proteins play a key role in cancer signaling and belong to the RAS superfamily.[22] We also identified a set of semaphorins in our analysis that is represented in the top significantly enriched pathways. Semaphorins, previously known as collapsins, are a set of proteins containing a 500-amino acid sema domain among others (including PSI and immunoglobulin type domains), which can be transmembranous or secreted.[23] It is known that Sema3E cleavage promotes invasive growth and metastasis *in vivo*.[23] These genes also have selected targeting by Rac and Rho family members. This generates hypotheses of possible pathways that could be targeted therapeutically.

## 4. Conclusions

We have introduced an integrated experimental and computational methodology that identifies the role of specific kinases in the drug response of a given cell line. The key element of our KIEN methodology is a multiple regression procedure that uses *in vitro* screen data as a training set. Were a new library of kinase inhibitor compounds be synthetized and profiled, our model would then be able to immediately estimate the effect of these drugs on *in vivo* experiments on a given cell line. We have shown an applications to a lung cancer cell line, but the method be extended to different cell lines. The method will facilitate the design of new kinase inhibitors and the development of therapeutic interventions with combinations of many inhibitors. The procedure could be extended to three drug combinations, if measurements for these larger combinations were available. Finally, the method could be extended to regression models that are specific of cancer cells with the same set of mutations, or it could be directly used with patient-derived primary cells to identify a personalized treatment.

## 5. Materials and Methods

### Materials

The primary screening of a kinase inhibitor (KI) library comprised of 244 KIs was purchased from EMD Chemicals, and diluted with DMSO to 2mM concentrations for high-throughput screening purposes. The KI library was stored at -80°C. Additionally, PDK1/Akt1/Flt3 Dual Pathway Inhibitor (CAS # 331253-86-2) was ordered from Calbiochem. Only 140 out of 244 were used in the drug-target network reconstruction because drug profiling information was available only for these compounds. One kinase inhibitor known to affect the kinase targets indirectly was excluded.

### Cell Culture

Cell lines IMR-90 (normal lung fibroblast) and A549 (lung adenocarcinoma) were cultured in RPMI 1640 (Hyclone) supplemented with 10% Canadian characterized fetal bovine serum (Hyclone), 1% 200mM L-glutamine (Omega), and 1% penicillin/streptomycin (Omega). The media for the cells were renewed every 3 days and kept at 80-90% confluency. Cells were maintained in a humidified environment at 37°C and 5% $CO_2$.

### Kinase Inhibitor Experiments

IMR-90 (1500 cells/well) and A549 (750 cells/well) were seeded on 384-well microplates (Grenier Bio-One) and incubated for 3 hours before the addition of kinase inhibitor(s). IMR-90 and A549 cell lines were tested on the same day with three replicates and the experiment was repeated three times with randomized well positions to reduce biases. ECHO 555 Liquid Handler (Labcyte) was used to dispense nanoliter volumes of each KI to 384-well plates with cells attached (wet dispense). The final volume in the plate is 40uL and cells were incubated for 72 hours with KI treatment.

*ATP Measurements*

ATPlite 1Step (Perkin Elmer) was used to evaluate the cell number and cytotoxicity. ATP measurements were done by dispensing 20uL of the ATPlite 1Step solution to each well with the final volume of 60uL. The plate was placed on a shaker at 1100rpm and luminescence activity was detected by Analyst GT Plate Reader. The percent (%) of control is the quantity of ATPlite 1step measurement of the treated versus the untreated wells of each individual cell type. The selectivity, S, is defined as *selectivity= (IMR-90/cell type)* as indicator of the kinase inhibitor's selective control of the kinase in the cancerous cell line (A549) over the normal cell line (IMR-90).

*Computational Methods*

Correlations between selectivity/viability and kinase activity were calculated using the python *scipy linregress* function, which also provide *p-values*. FDR were then obtained by ranking the *p-values* and directly applying the Benjamini–Hochberg procedure. The elastic net regression was carried out using the Scikit-learn package[24] which finds the coefficients $\beta$ that minimize the function

$$F = \frac{1}{2\,M}||v - A\beta||_2^2 + \alpha\rho||\beta||_1 + \frac{1}{2}\alpha(1-\rho)||\beta||_2^2 \, ,$$

where $v$ is the vector of the observed viabilities and $A$ is the matrix containing the residual activity of the kinases from the profiling. The parameters $\alpha$ and $\beta$ determine the relative weights of the lasso and ridge penalties quantified using $L^1$ ($||\cdot||_1$) and $L^2$ ($||\cdot||_2$) norm, respectively. We used $\alpha = 0.15$ and $\rho = 0.01$ in the results of Figures 4 and 5 and in Table 3. We also tried other values of these parameters, which did not give a significant difference in the results.

*Pathway-based enrichment*

Reactome pathways were downloaded using a newer build of the 'biomaRt' library (v2.12.0) in Bioconductor /R (v2.15.0). Gene symbols from the kinase list were converted to Entrez gene identifier numbers ('entrezgene') and mapped against the gene ids in each Reactome pathway. For each pathway, the set of significant genes enriched within any given pathway was computed using a Fisher exact test. The procedure computes the significance (p-value) of observing significant kinases, as deemed significant by our method, within the selected pathway. Given that our gene set consists entirely of kinases and would be generalized towards kinase-specific effects, the set of all kinases (~300) were selected for background adjustment and more sensitive enrichment of the pathways. This procedure was repeated for each pathway to generate p-values and pathway rankings. False discovery rate [FDR] values were later generated to further restrict significance.

The data sets supporting the results of this article are included within the article. Authors' contributions: GP & CP proposed concept, EO, CP and APH performed calculations, TT performed experiment, CP and GP wrote the manuscript. Correspondence and requests for materials should be addressed to giovanni@sanfordburnham.org or carlo@pa.msu.edu.

FIGURES

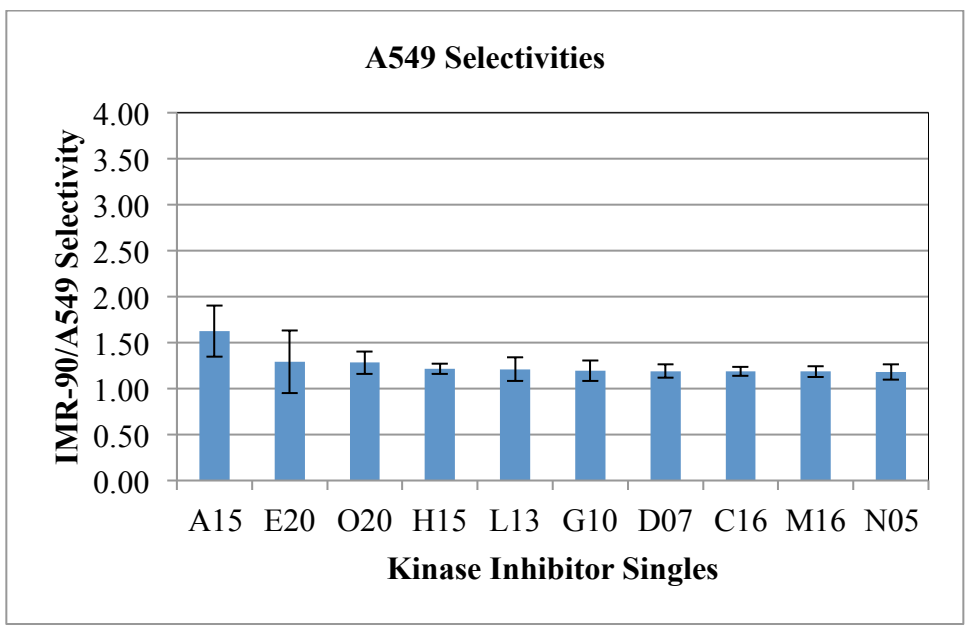

Figure 1. Primary screen results of the top ten most selective kinase inhibitors. The 3 digit code is the in-house code. A15: PDK1/Akt1/Flt3 Dual Pathway Inhibitor.

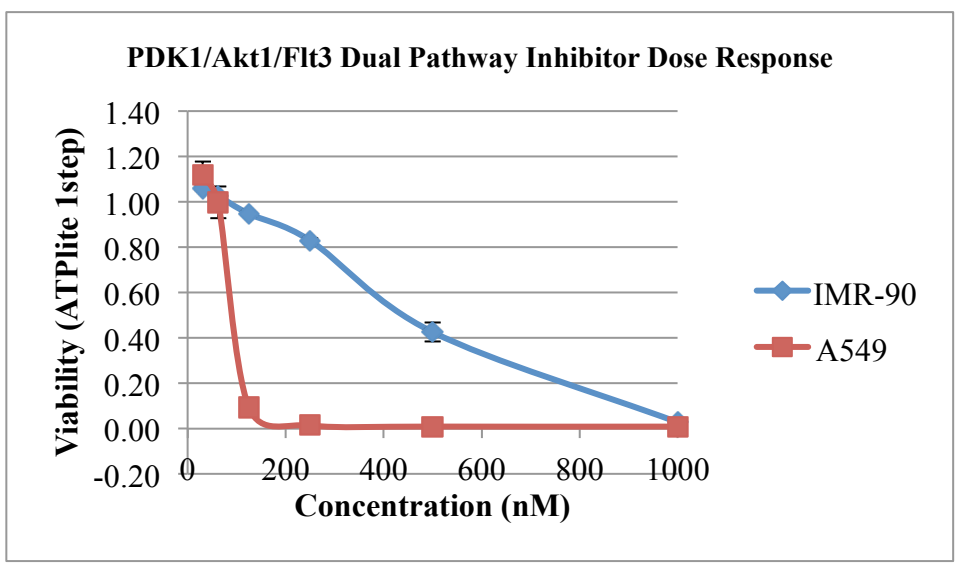

Figure 2. Optimization of PDK1/Akt1/Flt3 Dual Pathway Inhibitor. Lower doses of PDK1/Akt1/Flt3 Dual Pathway Inhibitor were tested to see the response of A549 to the drug. The doses are 31.25nM, 62.5nM, 125nM, 250nM, 500nM, and 1000nM.

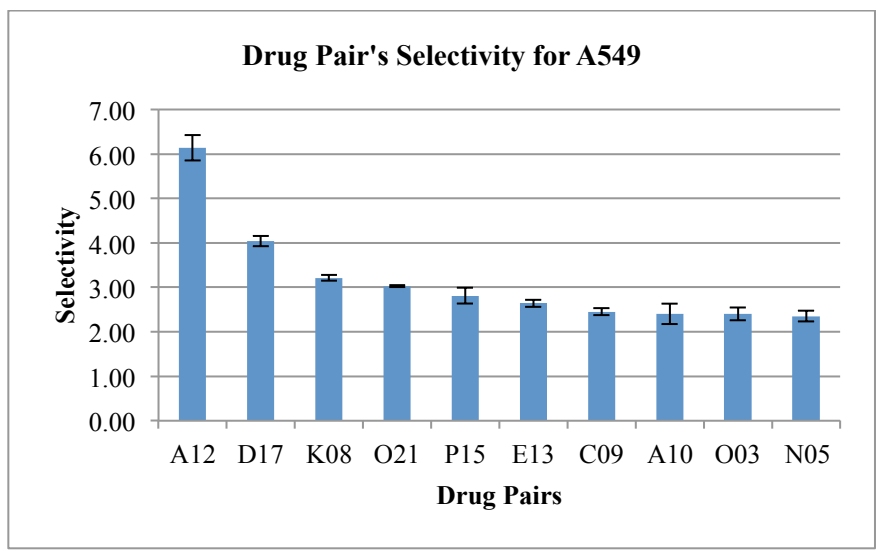

Figure 3. Secondary screen results of the top ten most selective drugs (1000nM) when paired with PDK1/Akt1/Flt3 Dual Pathway Inhibitor (125nM). The 3 digit code is the in-house code.

**Leave-one-out Cross Validation**

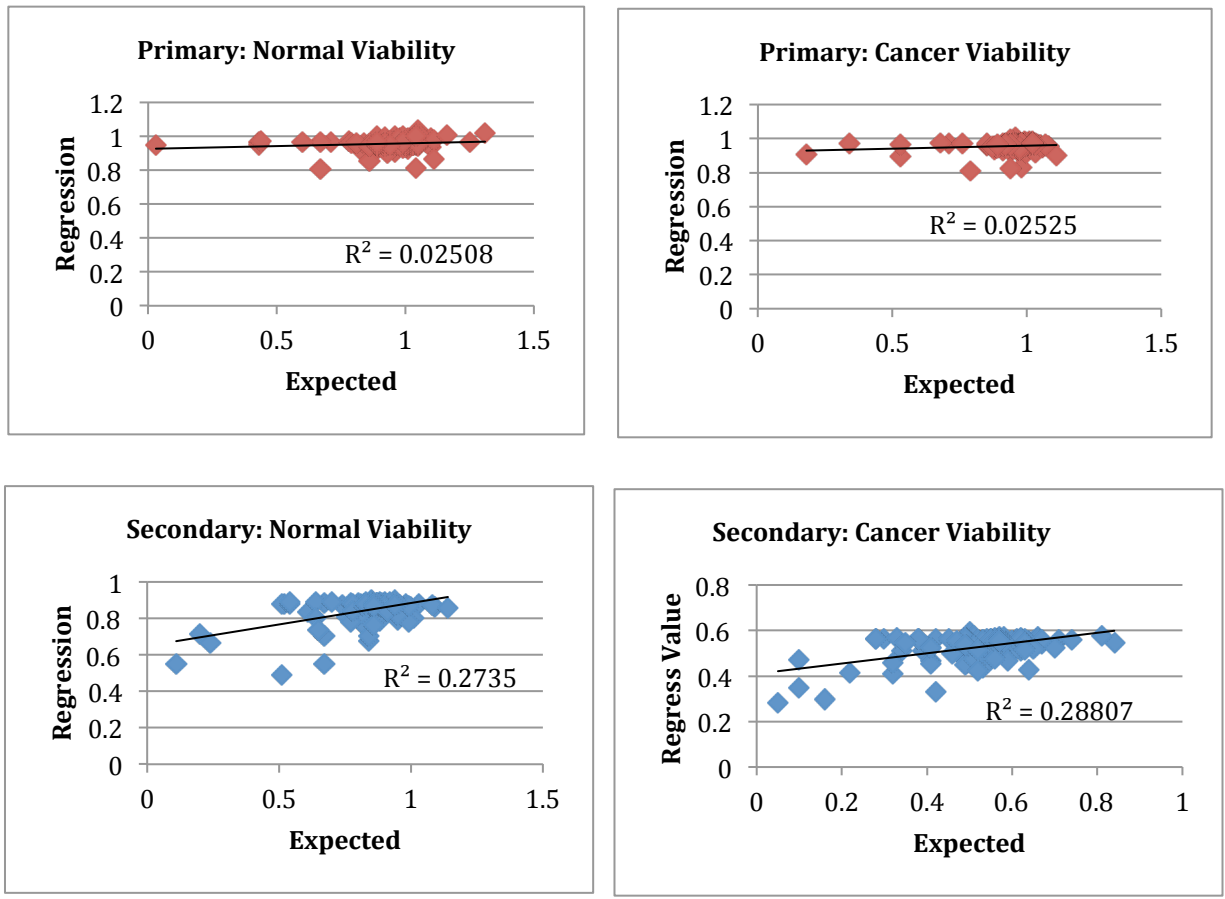

Figure 4: Leave-one-out Cross Validation of the elastic net regression model based on the primary (top) and secondary (bottom) screens for normal and cancer cell lines. Each of the 140 point in these figures corresponds to one of the 140 drug. "Regression" refers to the viability predicted by the regression model using all data from the other 139 drugs as training set, while "Expected" refers to the actual viability measured for the drug or drug combination. Note that only the secondary screen leads to predictive models with significant $R^2$ for the two cancer cell types.

**Leave one out cross validation: Log transformed data**

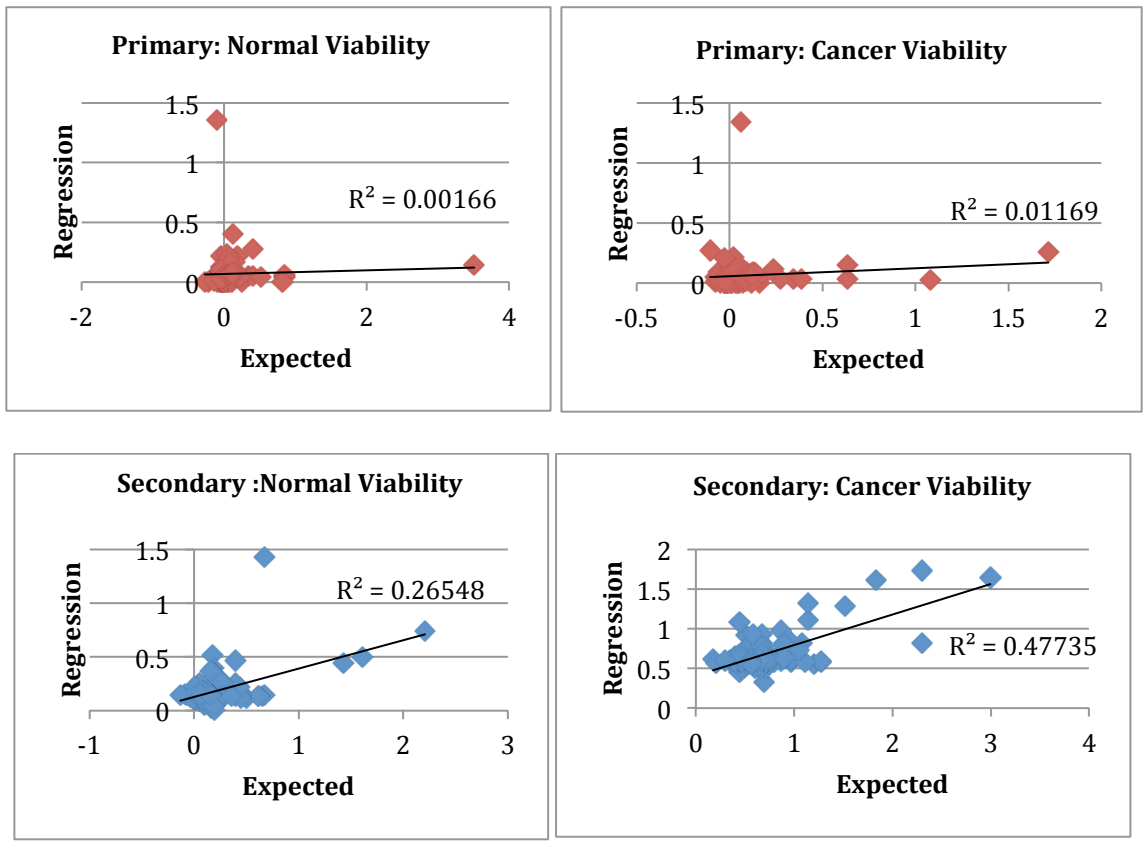

Figure 5: Leave-one-out Cross Validation of the elastic net regression model based on the primary (top) and secondary (bottom) screens for normal and cancer cell lines after logarithmic transformation on the data. Each of the 140 point in these figures corresponds to one of the 140 drug. "Regression" refers to $-log$ of the viability predicted by the regression model using all data from the other 139 drugs as training set, while "Expected" refers to $-log$ of the actual viability measured for the drug or drug combination. Note that, as in Figure 4, only the secondary screen leads to predictive models with significant $R^2$ for both cell types. The $R^2$ for the Cancer cell lines is considerably better using the log transformation.

TABLES

| Kinase | Selectivity Corr | FDR | Kinase | Selectivity Corr | FDR |
|--------|------------------|-----|--------|------------------|-----|
| Primary screening | | | Secondary Screening | | |
| PRKCZ | 0.451 | 2.28E-08 | TGFBR2 | -0.501 | 8.29E-08 |
| DMPK | 0.435 | 7.75E-08 | CDK4 | -0.412 | 6.40E-05 |
| STK39 | 0.430 | 1.15E-07 | CDC42BPB | -0.409 | 6.40E-05 |
| EPHA8 | 0.420 | 2.33E-07 | RIPK2 | -0.399 | 7.73E-05 |
| ADRBK2 | 0.399 | 1.01E-06 | DSTYK | -0.369 | 0.000413 |
| PRKACG | 0.396 | 1.27E-06 | ACVRL1 | -0.368 | 0.000413 |
| CAMK4 | 0.394 | 1.45E-06 | PAK1 | -0.367 | 0.000413 |
| MAP2K2 | 0.393 | 1.53E-06 | MAPKAPK2 | -0.364 | 0.000413 |
| ADRBK1 | 0.392 | 1.62E-06 | PAK7 | -0.359 | 0.000424 |
| PNCK | 0.382 | 3.29E-06 | CDK1 | -0.357 | 0.000429 |

Table 1 Correlations between selectivity and kinase activity from primary and secondary screening.

| Kinase | Normal Viab Corr | FDR | Kinase | Normal Viab Corr | FDR |
|---|---|---|---|---|---|
| Primary screening | | | Secondary Screening | | |
| ADRBK1 | 0.484 | 3.92E-07 | PAK1 | 0.653 | 6.23E-16 |
| DMPK | 0.481 | 3.92E-07 | PAK3 | 0.647 | 8.53E-16 |
| DDR2 | 0.459 | 1.15E-06 | PKN2 | 0.619 | 3.32E-14 |
| ZAP70 | 0.455 | 1.16E-06 | PDPK1 | 0.605 | 1.86E-13 |
| AKT2 | 0.453 | 1.16E-06 | SIK2 | 0.575 | 6.37E-12 |
| CAMK1G | 0.427 | 6.99E-06 | MAP3K10 | 0.568 | 1.22E-11 |
| TSSK2 | 0.428 | 6.99E-06 | FGFR2 | 0.562 | 1.98E-11 |
| MAPKAPK2 | 0.424 | 6.99E-06 | CAMK2G | 0.561 | 2.14E-11 |
| CAMK4 | 0.423 | 6.99E-06 | CDC42BPB | 0.559 | 2.20E-11 |
| PRKCZ | 0.419 | 7.54E-06 | NUAK2 | 0.545 | 9.97E-11 |

| Kinase | Cancer Viab Corr | FDR | Kinase | Cancer Viab Corr | FDR |
|---|---|---|---|---|---|
| Primary screening | | | Secondary Screening | | |
| AKT2 | 0.549 | 6.01E-10 | PAK1 | 0.625 | 4.51E-14 |
| DMPK | 0.545 | 6.01E-10 | PKN2 | 0.567 | 4.33E-11 |
| CAMK1G | 0.528 | 1.87E-09 | PAK3 | 0.554 | 1.25E-10 |
| DDR2 | 0.506 | 1.35E-08 | MAP3K10 | 0.550 | 1.43E-10 |
| ZAP70 | 0.493 | 3.67E-08 | CDC42BPB | 0.539 | 3.81E-10 |
| CAMK4 | 0.488 | 4.50E-08 | MAP3K2 | 0.520 | 2.16E-09 |
| ADRBK1 | 0.477 | 1.07E-07 | PDPK1 | 0.519 | 2.16E-09 |
| SGK3 | 0.477 | 1.07E-07 | SIK2 | 0.519 | 2.16E-09 |
| GRK5 | 0.473 | 1.20E-07 | NUAK2 | 0.502 | 8.14E-09 |
| AKT1 | 0.454 | 5.09E-07 | FES | 0.502 | 8.14E-09 |

Table 2 Correlation between viability of normal (top) and cancer (bottom) cell lines and kinase activity from primary and secondary screening.

| Kinase | Cancer beta coefficient | Normal beta coefficient | Difference |
|---|---|---|---|
| TGFBR2 | 0.061 | 0.000 | 0.061 |
| EGFR | 0.060 | 0.000 | 0.060 |
| PHKG1 | 0.051 | 0.014 | 0.037 |
| RIPK2 | 0.032 | -0.002 | 0.034 |
| PRKG2 | 0.012 | 0.045 | 0.033 |
| CDK4 | 0.021 | -0.008 | 0.029 |
| MAP3K10 | 0.038 | 0.014 | 0.024 |
| MARK4 | 0.000 | 0.022 | 0.022 |
| PAK1 | 0.025 | 0.004 | 0.021 |
| MAP4K5 | 0.021 | 0.000 | 0.021 |
| MARK2 | 0.006 | 0.026 | 0.021 |
| MARK3 | 0.000 | 0.020 | 0.020 |
| TBK1 | 0.012 | 0.031 | 0.020 |
| ERBB2 | 0.021 | 0.001 | 0.019 |
| NUAK1 | -0.029 | -0.010 | 0.019 |
| ULK2 | 0.018 | 0.000 | 0.018 |
| MYLK2 | -0.024 | -0.006 | 0.018 |
| MAP4K4 | 0.004 | -0.014 | 0.018 |
| CDK5 | 0.002 | -0.016 | 0.018 |
| GSK3B | 0.021 | 0.004 | 0.017 |
| PAK2 | 0.019 | 0.002 | 0.017 |
| CDC42BPB | 0.023 | 0.006 | 0.017 |
| DSTYK | 0.006 | -0.010 | 0.016 |
| RPS6KA2 | 0.000 | -0.016 | 0.016 |
| FGFR1 | -0.004 | 0.012 | 0.016 |
| PAK7 | 0.015 | 0.000 | 0.015 |
| PIM1 | -0.015 | 0.000 | 0.015 |
| CDK3 | 0.015 | 0.000 | 0.015 |
| IRAK1 | -0.002 | -0.017 | 0.015 |

Table 3. Kinases with the highest difference in the regression coefficients for the log transformed data of the secondary screen.

| Path ID | Path name | $N_S$ | $N_T$ | p-val |
|---------|-----------|-------|-------|-------|
| 422475 | Axon guidance | 8 | 31 | 0.006 |
| 428540 | Activation of Rac | 3 | 5 | 0.008 |
| 373755 | Semaphorin interactions | 4 | 10 | 0.011 |
| 376176 | Signaling by Robo receptor | 3 | 7 | 0.024 |
| 1266738 | Developmental Biology | 8 | 39 | 0.026 |
| 445144 | Signal transduction by L1 | 4 | 13 | 0.030 |
| 373760 | L1CAM interactions | 4 | 14 | 0.040 |
| 193639 | p75NTR signals via NF-kB | 2 | 4 | 0.051 |
| 209543 | p75NTR recruits signalling complexes | 2 | 4 | 0.051 |
| 389359 | CD28 dependent Vav1 pathway | 2 | 4 | 0.051 |

Table 4. Reactome pathways with significant representation of kinases from the regression analysis. $N_s$ indicates the number of kinases that are found significant in the regression analysis, while $N_T$ is the total number of kinases in the pathway. The top nine pathways with Fisher exact test p<0.051 are shown. These pathways are identified from 518 Reactome pathways containing at least one of the kinases identified in Table 3.